\documentstyle[12pt]{article}
\parskip=\medskipamount                 
\thicklines                       
\begin{document}
\titlepage
\begin{flushright}
{IMSc--92/35}\\
{hepth@xxx/9208064}
\end{flushright}
\begin{center}
{\bf GENUS ZERO CORRELATION FUNCTIONS IN  $c<1$ STRING THEORY}\\
Suresh Govindarajan, T. Jayaraman and Varghese John\footnote{email:
suresh, jayaram, john@imsc.ernet.in}\\
The Institute of Mathematical Sciences\\
C.I.T. Campus, Taramani\\
Madras 600113 INDIA\\
\bigskip
\bigskip
\bigskip
{\bf Abstract}
\end{center}
\bigskip
We compute N-point
correlation functions of pure vertex operator states(DK states)
for minimal models coupled to gravity.
We obtain agreement with the matrix model results on
analytically continuing in the numbers of cosmological constant
operators and matter screening operators. We
illustrate this for the cases of the $(2k-1,2)$ and $(p+1,p)$ models.
\vfil
\begin{flushleft}
August, 1992
\end{flushleft}
\newpage
\section{Introduction}
One of the important issues in the understanding of $c<1$ models coupled
to 2-d gravity has been the computation of the correlation functions in
the continuum approach. These are well understood from the matrix model
point of view. The most convenient method to calculate such correlation
functions in the continuum approach is of course to use the free field
representation of the $c<1$ models.
Three point functions on the sphere for states within the conformal
grid were first calculated by this method in ref. \cite{GLI} and
then extended to 3-point functions with pure vertex operator states(DK
states) with momenta outside the Kac table\cite{KIT,DOT,SWA}. These
calculations
matched the matrix model results. N-point correlation
functions(of states inside the Kac table) that did not need matter screening
were calculated in ref.\cite{DFrK}.
In particular the role of the screening operator was not clear.
In this paper, we calculate arbitrary N-point
correlation functions(with matter screening) involving DK states for
the cases of $(2k-1,2)$
and $(p+1,p)$ models. The general case will be dealt elsewhere.
We shall show that these results agree with the
corresponding matrix model results. This agreement requires an
analytic continuation in the number of screening operators just as in
refs. \cite{GLI,KIT,KAED}.
The free field representation of the matter part of the theory,
involves the use of a screening operator as a BRST operator\cite{GF}.
In the case of matter coupled to gravity, the physical states are
given by the double cohomology of this BRST operator and the usual
string BRST.

It was shown that DK states are also physical and are related to
physical states of higher ghost
number by means of descent equations in the two BRST operators\cite{descent}.
It was also shown
how correlation functions with states of higher ghost number could be
converted into correlation functions involving DK states. Hence,
calculation of correlation functions with DK states should provide a
complete description of the theory. There is also a ring of ghost number
zero operators in the theory\cite{ring} (similar to those found for
$c=1$ coupled to Liouville\cite{Witten}).
These operators prove useful in obtaining
recursion relations involving DK states and enable us to convert
integrals which are quite hard to compute into
those for which the results of Dotsenko and Fateev\cite{DF} are sufficient.
This enables us to calculate all correlation functions which require positive
number of screening in the Liouville and matter sector. We then obtain
all other correlation functions by analytically continuing these
results to fractional number of screening operators.

\section{Classification of DK States}
The energy-momentum tensors of the matter and Liouville fields are
given by
\begin{eqnarray*}
T^M &= -\frac14\partial X \partial X  + i \alpha_0 \partial^2X\quad,\\
T^L &= -\frac14\partial \phi \partial \phi  + i \beta_0 \partial^2\phi\quad,
\end{eqnarray*}
with central charges $c_M=1-24\alpha_0^2$ and $c_L=1-24\beta_0^2$. For the
$(p',p)$ model, $2\alpha_0=\frac{(p'-p)}{\sqrt{p'p}}$ and $2\beta_0=
\frac{i(p'+p)}{\sqrt{p'p}}$. The physical vertex operators are of the
form  $exp(i\alpha X + i\beta \phi)$.
In the matter part, there are two screening operators $Q_\pm$
$$
Q_+ = \Delta(1+\frac{p'}p)\int e^{i\alpha_+X},\quad
Q_- = \Delta(1+\frac{p}{p'})\int e^{i\alpha_-X}
$$
where $
\alpha_+ = {{p'} \over {\sqrt{pp'}}},\quad
\alpha_- = {{-p} \over {\sqrt{pp'}}}$ and $\Delta(x)\equiv
\frac{\Gamma(x)}{\Gamma(1-x)}$.
Note that the measure factor of ${\frac{d^2z}{\pi}}$
is implicit in all integrals.
Both the screening operators furnish equally good BRST operators(in
the sense of Felder\cite{GF}). One finds that
physical operators can be represented by
vertex operator states(DK states) with matter momenta both inside and
outside the Kac table\cite{DOT,KIT,descent}. However,  pure vertex
operator states from outside the Kac table depend on the choice of
screening operator as a BRST operator(we call this a choice of
resolution). We shall deal only with operators
which satisfy the condition $\beta<\beta_0$ in order that their
scaling dimensions match with those obtained in the matrix models.
This(choice of resolution) leads to two distinct sets
of DK states. Further, one
has to complete the set by including states belonging to the edge of
the Kac table.
For the
case of the $(p+1,p)$, the DK states in the $Q_-$ resolution are
\begin{equation}
V_n^\alpha = exp{{[p(n-2)+\alpha]\phi + [pn+\alpha+2]iX
 } \over {2\sqrt{p(p+1)}}}\quad,
\end{equation}
where $\alpha=0,\ldots,(p-2)$. This includes edge states of the form
$(j(p+1),m)$. We have excluded the ``wrong-edge'' states which correspond
to $\alpha=(p-1)$\cite{ring}.  $V_0^0$ is the cosmological constant operator.
Also, the other screening operator $Q_+$($\sim\int V_2^0$) is now
a physical operator. Thus $Q_+$ cannot be used as a screening operator
anymore. For momentum conservation in the matter sector, we shall
restrict to using $Q_-$ exclusively as a screening operator. Hence, we
will need to use analytic continuation in the matter sector when necessary.

Similarly, the DK states in the $Q_+$ resolution are
\begin{equation}
V_n^\alpha = exp{{[(p+1)(n-2)+\alpha+2]\phi + [-(p+1)n-\alpha]iX}
 \over {2\sqrt{p(p+1)}}}
\end{equation}
where $\alpha=0,\ldots,(p-1)$. This includes edge states of the form
$(m',jp)$. The other screening operator $Q_-$ is now a physical operator.
We normalise the pure vertex operators by
$\Delta(n + {{(\alpha+1)}\over p})$ for the $Q_-$ resolution and
$\Delta(n + {{(\alpha+1)}\over {(p+1)}})$ for the $Q_+$
resolution.
Note that this normalisation is not singular for the allowed values of
$\alpha$.

\section{Correlation Functions of DK states}
\subsection{$(2k-1,2)$ Models}
The DK states for the $(2k-1,2)$ models in the $Q_-$ resolution are given by
\begin{equation}
V_n = \Delta(1+n+{1\over2})~~ exp {{(n-k)\phi + (n+k-1)iX} \over
{\sqrt{2(2k-1)}}}\quad,
\end{equation}
where $n=0,1,\ldots$. Note that the ``physical'' screening operator
$Q_+$ is given by $V_k$.
The ring elements are generated by\cite{ring}
\begin{equation}
a_- = -|bc + {1\over2}\sqrt{{(2k-1)}\over{2}}\partial(\phi - iX)|^2
{}~~exp{{(\phi + iX)}\over {\sqrt{2(2k-1)}}}\quad.\label{eringt}
\end{equation}
The ring elements are $(a_-)^n$. The action of the ring element on the
DK states is given by\cite{Kachru,KMS,bershad}
\begin{equation}
\lim_{z\rightarrow w}a_-(z) c{\bar c}V_n(w) \sim c{\bar c}V_{n+1}(w),
\quad a_-(z) c{\bar c}e^{i\alpha_-X}(w)\sim 0
\end{equation}
Now consider a charge conserving correlation function with one $a_-$
and DK states

\begin{equation}
F(w,{\bar w}) \equiv \langle a_-(w) c{\bar c}V_{n_1}(0) c{\bar c}V_{n_2}(1)
c{\bar c}V_{n_3}(\infty) \prod_{i=4}^{N} \int
V_{n_i} {1\over{R!}}(Q_-)^R\rangle
\end{equation}

One can check that $\partial_w F=\partial_{\bar w} F=0$ using
$\partial_w a_- =\{Q_B, b_{-1}a_-\}$ and then deforming the contour of
$Q_B$. This implies that $F(w,{\bar w})$ is a constant independent of
$w$(${\bar w}$). Equating $F(0)$ with $F(1)$ and using
eqn.(\ref{eringt}), we obtain
\begin {eqnarray}
\langle  c{\bar c} V_{n_1+1}(0) c{\bar c} V_{n_2}(1)
c{\bar c} V_{n_3}(\infty) \prod_{i=4}^{N} \int
 V_{n_i} {1\over{R!}} (Q_-)^R\rangle \nonumber \\
 =\langle  c{\bar c} V_{n_1}(0) c{\bar c} V_{n_2+1}(1)
c{\bar c} V_{n_3}(\infty) \prod_{i=4}^{N}  \int
 V_{n_i}{1\over{R!}}(Q_-)^R\rangle\quad.
\end{eqnarray}
This gives us the following recursion relation(similar to the one in
ref. \cite{distler})
\begin{equation}
V_{n+1}(z)V_m(w)
= V_{n}(z)V_{m+1}(w)  \label{recur}
\end{equation}
In particular, $V_nV_0=V_{n-k}V_k$. This is sufficient to convert all
correlation functions(with positive integer number of Liouville
screening)  to ones which are of the Dotsenko-Fateev type\cite{DF}
making them computable. We shall now demonstrate this
explicitly. Consider the correlation function
\begin{equation}
\langle \langle \prod_{i=1}^{L} \int V_{n_i} \rangle \rangle =
\mu^S \Gamma(-S) \langle c{\bar c}V_{n_1}(0)c{\bar c}V_{n_2}(1)
c{\bar c}V_{n_3}(\infty)
\prod_{i=4}^{L} \int V_{n_i} (\int V_0)^S
{1\over{R!}}
(Q_-)^R \rangle ,
\end{equation}
where $S= (2-L) + {1\over k}{(\sum_{i=1}^L n_i) +{1\over k}}$
and $2R=2 (\sum_{i=1}^L n_i)- (L+S-4)$. The expressions for $S$ and
$R$ follow from the charge conservation relations in Liouville and
matter respectively. The RHS of the above equation is obtained after
completing the zero-mode integration of the Liouville mode\cite{GLI}
and introducing screening in the matter sector for charge conservation.
We shall assume that both $R$ and $S$ are positive
integers. We can now use the recursion relation, to obtain
\begin{equation}
\langle \langle \prod_{i=1}^{L} \int  V_{n_i} \rangle \rangle
=\mu^S \Gamma(-S)\langle c{\bar c} V_0(0) c{\bar c} V_0(1)
c{\bar c} V_{k-1}(\infty) (\int V_{k})^{L+S-3}
{1\over{R!}} (Q_-)^R \rangle
\end{equation}
This correlator can be evaluated using formula (B.10) of
ref.\cite{DF}. This gives
$$
\langle \langle \prod_{i=1}^{L} \int  V_{n_i} \rangle \rangle
= \mu^S \Gamma(-S) (L+S-3)! {{\Gamma(1)}\over{\Gamma(0)}}  \rho
=\mu^S{{\Gamma(L+S-2)}\over{\Gamma{(S+1)}}}\rho\quad,
$$
where $\rho = {2\over{(2k-1)}}$.
The cases of non-integer $S$ and $R$ are obtained by analytically
continuing the above result to non-integer values. Hence, one obtains
\begin{equation}
\langle \langle \prod_{i=1}^{L} \int V_{n_i} \rangle \rangle =
\rho \mu^S {{\Gamma(\frac{(\sum_in_i) +1}{k})}\over {\Gamma(S+1)}}\quad,
\end{equation}
This result is in complete agreement with matrix model
results\cite{matrix,distler}. This
represents the generalisation of the 3 point function calculations
given in \cite{GLI,KIT}. The use of the recursion relation has
facilitated the calculation of arbitrary N-point functions.

\subsection{$(p+1,p)$ models}
The DK states in the $Q_-$ resolution are
$$
V_n^\alpha = \Delta(n + {{(\alpha+1)}\over p})exp{{[p(n-2)+\alpha]\phi
+ [pn+\alpha+2]iX } \over {2\sqrt{p(p+1)}}}
$$
where $\alpha=0,\ldots,(p-2)$.
The ring elements for these models are generated by\cite{KMS}
\begin{eqnarray}
a_+&=  -|bc + {1\over2}\sqrt{{p}\over{p+1}}\partial(\phi + iX)|^2
{}~~exp{{(p+1)(\phi - iX)}\over
{2\sqrt{p(p+1)}}},\nonumber \\
a_- &= -|bc + {1\over2}\sqrt{{p+1}\over{p}}\partial(\phi - iX)|^2
{}~~ exp{{p(\phi + iX)}\over {2\sqrt{p(p+1)}}}.
\end{eqnarray}
The ring elements in the $Q_-$ resolution are\cite{ring}
\begin{equation}
(a_-)^n,~~ a_+(a_-)^n,~~ \ldots,~~(a_+)^{p-1}(a_-)^n\quad. \nonumber
\end{equation}
with the equivalence relation $a_+^p\sim a_-^{p+1}$.
The action of the ring on DK states is\cite{Kachru,KMS,bershad}
\begin{eqnarray}
\lim_{w\rightarrow z}a_-(w)c{\bar c}V_n^\alpha(z) \sim c{\bar c}
V_{n+1}^\alpha(z)\quad,\nonumber\\
\lim_{w\rightarrow z}a_+(w)c{\bar c}V_n^\alpha(z)\int V_m^\beta(t)
\sim c{\bar c} V_{n+m-1}^{\alpha+\beta +1}(z)\quad,
\label{eringb}
\end{eqnarray}
where we have normalised all the DK states by their appropriate
leg-factors. We shall now use the ring elements to calculate arbitrary
N-point functions explicitly for the case of the Ising model which is
the  $(4,3)$ model. We shall then present results for the general
$(p+1,p)$ models. Details of the calculations will be given elsewhere.
The case of pure gravity i.e., the $(3,2)$ model has already been
obtained in the earlier section. In the Ising model, $V_0^0$
corresponds to the identity operator, $V_0^1=\sigma$ and
$V_1^1=\epsilon$ correspond to the spin and energy operators
respectively(with appropriate Liouville dressing). $V_2^0$ is the
``physical'' screening operator.
One can prove the following
shift recursion relation. The arguments are identical to the one used
in deriving eqn.(\ref{recur}).
\begin{equation}
V_{n+1}^\alpha(z)V_m^\beta(w)
= V_{n}^\alpha(z)V_{m+1}^\beta(w)\label{erecurb}
\end{equation}
Now consider a charge conserving correlation function with one $a_+$
and DK states.
\begin{equation}
\langle a_+(w) c{\bar c}V_{n}^0(0) c{\bar c}V_{m}^1(1)
c{\bar c}V_{r}^\alpha(\infty)
\prod_{i=1}^{L} \int V_{n_i}^0
\prod_{j=1}^{M} \int V_{m_i}^1
 {1\over{R!}} (Q_-)^R\rangle
\end{equation}
Using the $w$ independence of the above correlation function, we
equate the value of the correlator at $w=0$ and $1$.
This gives us after using the second equation in (\ref{eringb})
\begin{eqnarray}
\sum_{k=1}^L \langle  c{\bar c}V_{n+n_k-1}^1(0) c{\bar c}V_{m}^1(1)
c{\bar c}V_{r}^\alpha(\infty)
\prod_{i=1,i\neq k}^{L} \int V_{n_i}^0
\prod_{j=1}^{M} \int V_{m_i}^1
\rangle \nonumber \\
+
\sum_{k=1}^M\langle  c{\bar c}V_{n+m_k-1}^2(0) c{\bar c}V_{m}^1(1)
c{\bar c}V_{r}^\alpha(\infty)
\prod_{i=1}^{L} \int V_{n_i}^0
\prod_{j=1,j\neq k}^{M} \int V_{m_i}^1
\rangle \nonumber \\
=\sum_{k=1}^{L}
\langle  c{\bar c}V_{n}^0(0) c{\bar c}V_{m+n_i-1}^2(1)
c{\bar c}V_{r}^\alpha(\infty)
\prod_{i=1,i\neq k}^{L} \int V_{n_i}^0
\prod_{j=1}^{M} \int V_{m_i}^1
\rangle \nonumber \\
+ \sum_{k=1}^{M}
\langle  c{\bar c}V_{n}^0(0) c{\bar c}V_{m+m_k}^0(1)
c{\bar c}V_{r}^\alpha(\infty)
\prod_{i=1}^{L} \int V_{n_i}^0
\prod_{j=1,j\neq k}^{M} \int V_{m_i}^1
\rangle
\end{eqnarray}
Using eqn. (\ref{erecurb}), one can see that every term inside each of the
sums are the same. One can now see that the following operator
relations provides a consistent solution to the above equality.
\begin{eqnarray}
V^2_m(z) V_n^1(w) &=&V_m^0(z)V_{n+1}^0(w)\nonumber\\
V^2_m(z) V_n^0(w)  &=& V_m^1(z)V_n^1(w)\label{erecurc}
\end{eqnarray}
However, $V^2_m$ belongs to the ``wrong-edge'' and is not
physical. However by using this relation twice, we obtain the following
recursion which involves only physical operators
\begin{equation}
V_m^1(z)V_n^1(w)V_r^1(t)=
V_{m+1}^0(z)V_n^0(w)V_r^0(t)\quad.\label{ewrecur}
\end{equation}
For the case of the $(4,3)$ model, eqns. (\ref{erecurb}) and
(\ref{ewrecur}) are sufficient to convert all charge conserving
correlation functions to those which are of the Dotsenko-Fateev type
and hence are computable. We shall now demonstrate this.
Consider the following correlation function
\begin{eqnarray}
\langle \langle \prod_{i=1}^L \int V_{n_i}^{\alpha_i} \rangle\rangle
&\nonumber \\
= \mu^S \Gamma(-S)& \langle c{\bar c}V_{n_1}^{\alpha_1}(0)c{\bar c}
V_{n_2}^{\alpha_2}(1)~c{\bar c}V_{n_3}^{\alpha_3}(\infty)
\prod_{i=4}^L \int V_{n_i}^{\alpha_i} (\int V_0^0)^S
{1\over {R!}} (Q_-)^R \rangle \quad,\label{eisinga}
\end{eqnarray}
where $S={{\sum_{i=1} n_i}\over 2} -L +2+ {{(\sum_{i=1}
\alpha_i)+2}\over 6}$
and $R=\sum_{i=1} ({{n_i}\over
2}+{{\alpha_i}\over 6}) +{{S}\over 3}$. When, $S$
and $R$ are positive integers, using  eqns. (\ref{erecurb})
and (\ref{ewrecur}) in eqn.(\ref{eisinga}) we obtain
\begin{equation}
\langle \langle \prod_{i=1}^L \int V_{n_i}^{\alpha_i} \rangle\rangle
=\mu^S\Gamma(-S) \langle c{\bar c}V_0^0(0)c{\bar c}V_0^0(1)~c{\bar
c}V_1^1(\infty)
(\int V_2^0)^{(L+S-3)}{1\over{R!}} (Q_-)^R \rangle \quad.\label{eisingb}
\end{equation}

The correlation function in the RHS can be explicitly computed using
the formula of Dotsenko and Fateev. We obtain
\begin{equation}
\langle \langle \prod_{i=1}^L \int V_{n_i}^{\alpha_i} \rangle\rangle
= \mu^S {{(L+S-3)!} \over {S!}} \rho\quad,\label{esingc}
\end{equation}
where $\rho={3\over4}$. This is in agreement with matrix model
results. For the cases
when $S$ and $R$ are not positive integers, the results are obtained
by analytic continuation. Of course, one has to take care that the
$Z_2$ invariance of the minimal models is not violated. One imposes
this by setting all non-$Z_2$ invariant correlators to zero by hand.
For the example of Ising model, the $Z_2$ charge of the operator
$V_n^\alpha$ is $(-1)^{n+\alpha}$. So the correlation function in
({\ref{eisinga}) is non-zero provided $\sum_i (n_i +\alpha_i)$ is
even. For such cases, the result one obtains after analytic
continuation in both $S$ and $R$ is
\begin{equation}
\langle \langle \prod_{i=1}^L \int V_{n_i}^{\alpha_i} \rangle\rangle
=\rho \mu^S
{{\Gamma( {{\sum_i n_i}\over 2}+ {{(\sum_i\alpha_i)+2}\over 6})}
\over {\Gamma{(S+1)}}}\quad.
\end{equation}
This generalises for the $(p+1,p)$ model. For  $Z_2$
invariant correlation functions one obtains
\begin{equation}
\langle \langle \prod_{i=1}^L V_{n_i}^{\alpha_i} \rangle\rangle
=\rho \mu^S
{{\Gamma( {{\sum_i n_i}\over 2}+ {{(\sum_i\alpha_i)+2}\over {2p}})}
\over {\Gamma{(S+1)}}}\quad,
\end{equation}
where $S={{\sum_{i=1} n_i}\over 2} -L +2+ {{(\sum_{i=1}
\alpha_i)+2}\over {2p}}$ and $\rho={p\over{p+1}}$.
(Note that the $Z_2$ charge of the operator $V_n^\alpha$ is
$(-1)^{(np+\alpha)}$ ).
\section{Discussion}

It is known that the DK states  and ring elements of $c<1$
string theory can be obtained by a target space Lorentz transformation
of $c=1$ string tachyons and ring elements\cite{ring}.
The correlation functions we have evaluated here
lead to integrals which are identical to those obtained in $c=1$
string theory provided one chooses tachyons of specific momenta and
chirality. Indeed, all DK states of a given resolution map on to
tachyons of the same chirality while the screening operator is of the
opposite chirality. However, the ${{\Gamma(1)}\over{\Gamma(0)}}$
factor is compensated for by the $\Gamma(-S)$ coming from the
zero-mode integration of the Liouville field. Evaluating the
integrals, in particular requires some care. The consistent
prescription appears to be the one where using the recursion relations
the integrals appear in a form such that formula (B.10) of \cite{DF}
is applicable. We also note that even for 3-point functions, it seems
that the Liouville and matter sectors cannot be integrated separately
as the results do not agree with the recursion relations.

The analytic continuation in the number of matter screening operators
and the cosmological constant operators as used here appears essential
to reproduce the matrix model results. The use of other physical operators as
screening operators\cite{ring,dot} does not appear to be possible
beyond three point functions.
Analytic continuation in matter may seem surprising but is
essential especially in the case of non-unitary models where the
cosmological constant operator carries non-trivial matter momentum.
An analytic continuation in the number of cosmological constant
operators in these models forces an analytic continuation in the
matter sector too. However, a deeper understanding of the origin of
this analytic continuation still eludes us.

The well known $\alpha\rightarrow(2\alpha_0-\alpha)$ duality symmetry
of the free field representation of minimal models has to be given up
for states outside the Kac table in order to be consistent with the
choice of resolution. Though this is not true for states inside the
Kac table, it is simpler to work with only the matter momenta given by
the parametrisation of the resolution.

The string equation of the $(p+1,p)$ model in the matrix model
approach can be obtained either from a $p-th$ order differential
operator or a $(p+1)-th$ order differential operator. This ambiguity
is reflected in the two possible parametrisations of the physical
states in the continuum.

We thus appear to have a complete prescription to calculate
correlation functions on the sphere for the arbitrary $(p',p)$ minimal
model coupled to gravity. It would be interesting to extend these
techniques to higher genus where again the matrix model results are
available.

{\flushleft{\bf Acknowledgements}}

We would like to thank P. Majumdar and P. Durganandini  for useful
discussions.


\begin{thebibliography}{99}
\bibitem{GLI}{ M. Goulian and B.Li,  Phys. Rev. Lett. {\bf 66}, 2051(1991).}
\bibitem{KIT}{ Y. Kitazawa, Phys. Lett. {\bf B265}, 262(1991).}
\bibitem{DOT}{ V. Dotsenko, Mod. Phys. Lett. {\bf A6}, 3601(1991).}
\bibitem{SWA}{D. Ghoshal and S. Mahapatra, ``Three-Point Functions of
Non-Unitary Minimal Matter coupled to gravity,'' Tata Preprint
TIFR/TH/91-58(1991).}
\bibitem{DFrK}{ P. Di Francesco and D. Kutasov, Nuc. Phys. {\bf B375},
119(1992).}
\bibitem{GF}{ G. Felder, Nucl. Phys. {\bf B317} (1989) 215.}
\bibitem{KAED}{ K. Aoki and E. D' Hoker, Mod. Phys. Lett.{\bf A7} 235(1992).}
\bibitem{descent}{ S. Govindarajan, T. Jayaraman, V. John and P.
Majumdar, Mod. Phys. Lett. {\bf A7}, 1063(1992).}
\bibitem{ring}{ S. Govindarajan, T. Jayaraman, and V. John, ``Chiral
Rings and Physical States in $c<1$ String Theory,'' IMSc
Preprint--92/30 = hepth@xxx/9207109.}
\bibitem{Witten}{ E. Witten, Nuc. Phys. {\bf B373}, 187(1992).}
\bibitem{DF}{ V. Dotsenko and Fateev, Nuc. Phys. {B251}, 691(1985).}
\bibitem{Kachru}{ S. Kachru, Mod. Phys. Lett. {\bf A7} 1419(1992).}
\bibitem{KMS} D. Kutasov, E. Martinec and N. Seiberg, Phys. Lett. {\bf
B276} 437(1992).
\bibitem{bershad}{M. Bershadsky and D. Kutasov, ``Scattering of Open
and Closed Strings in $1+1$ dimensions,'' Preprint PUPT-1315 =
HUTP-92/A016 (1992).}
\bibitem{distler}{ J. Distler, Nuc. Phys. {\bf B342}, 523(1990).}
\bibitem{matrix}{D. Gross and A. Migdal, Phys. Rev. Lett.{\bf 64} 127(1990).
R. Dikjgraaf and E. Witten, Nuc. Phys. {\bf B342} 486(1990).}
\bibitem{dot}{Vl. S. Dotsenko, ``Correlation Functions of Local
Operators in 2D gravity coupled to minimal matter,'' Preprint
PAR-LPTHE 91-52(1991).}
\end{thebibliography}
\end{document}